\documentclass[twocolumn,english]{revtex4-1}
\usepackage[T1]{fontenc}
\usepackage[a4paper]{geometry}
\usepackage{hyperref}
\usepackage{babel}
\usepackage{amsmath}
\usepackage{tikz}

\begin{document}

\title{New recursions for tree-level correlators in (Anti) de Sitter space}

\author{Connor Armstrong$^{a}$}
\author{Humberto Gomez$^{a,b}$}
\author{Renann Lipinski Jusinskas$^{c}$}
\author{Arthur Lipstein$^{a}$}
\author{Jiajie Mei$^{a}$}
\affiliation{$^{a}$ Department of Mathematical Sciences, Durham University, ~\\
Stockton Road, DH1 3LE Durham, UK}
\affiliation{$^{b}$ Facultad de Ciencias Basicas, Universidad Santiago de Cali,~\\
Calle 5 $N^\circ$ 62-00 Barrio Pampalinda, Cali, Valle, Colombia}
\affiliation{$^{c}$ Institute of Physics of the Czech Academy of Sciences \& CEICO
~\\
Na Slovance 2, 18221 Prague, Czech Republic}

\begin{abstract}
We present for the first time classical multiparticle solutions in Anti de Sitter space (AdS) involving scalars, gluons, and gravitons. They are recursively defined through multiparticle currents which reduce to Berends-Giele currents in the flat space limit. This construction exposes a compact definition of tree-level boundary correlators using a general prescription that removes unphysical boundary contributions.  Similarly to the flat space perturbiner, a convenient gauge choice leads to a scalar basis for all degrees of freedom, while the tensor structure is exclusively captured by field theory vertices. This provides a fully automated way to compute AdS boundary correlators to any multiplicity and cosmological wavefunction coefficients after Wick-rotating to de Sitter space. 
\end{abstract}

\maketitle

\section{Introduction}

Quantum field theory (QFT) in curved spacetime is full of subtleties. For example, it is possible to construct an S-matrix in asymptotically flat backgrounds, but its definition is unclear in generic spacetimes. Thus, general techniques that we can export from flat to curved spaces are very welcome. Naturally, studying QFT in curved spacetime is also relevant for understanding quantum gravity. Of particular interest are backgrounds with non-zero cosmological constant, notably (Anti) de Sitter space -- (A)dS. Boundary correlators in these spacetimes are constrained by conformal Ward identities and reduce to bulk scattering amplitudes in the flat space limit. In AdS, this underlies the gauge-gravity duality between conformal field theory (CFT) and string theory \cite{Maldacena:1997re}. In dS, this provides a powerful new set of tools for computing cosmological observables inspired by scattering amplitudes, which is now a very active area of research (see e.g. \cite{Baumann:2022jpr} for a recent review). 

Recursive techniques have had a major impact on the understanding of flat space scattering amplitudes, and are therefore valuable goals to pursue in curved space. For instance, the BCFW recursion \cite{Britto:2005fq} was generalized to AdS in \cite{Raju:2010by,Raju:2011mp,Raju:2012zr}, while recursions for Witten diagrams were developed in \cite{Arkani-Hamed:2017fdk,Yuan:2017vgp,Zhou:2018sfz} for scalars, and more generally in \cite{Zhou:2020ptb}. However, they do not exhibit the same level of efficiency as flat space recursions and cannot be directly used to compute correlators involving more than one type of particle. Alternatively, the Berends-Giele (BG) recursion \cite{Berends:1987me} (later extended and formalized in \cite{Kim:1996nd}) provides a clearer path for the computation of curved space correlators. In flat space, BG currents can be seen as tree-level amplitudes with one off-shell leg. Higher-point amplitudes are then built by connecting BG currents through field theory vertices. More recently, the BG recursion was partially extended to AdS embedding space \cite{Herderschee:2022ntr,Cheung:2022pdk}, yielding a differential representation for boundary correlators with external scalars, although a practical extension to spinning particles remained elusive.

In this letter we take this extra step and establish the AdS generalization of the so-called perturbiner method \cite{Rosly:1996vr,Rosly:1997ap,Selivanov:1998hn} (see also \cite{Mafra:2015gia,Lee:2015upy,Mafra:2015vca,Mafra:2016ltu,Mafra:2016mcc,Mizera:2018jbh,Garozzo:2018uzj,Lopez-Arcos:2019hvg,Gomez:2020vat,Guillen:2021mwp,Gomez:2021shh,Cho:2021nim,Ben-Shahar:2021doh,Escudero:2022zdz,Lee:2022aiu,Gomez:2022dzk} for a number of recent applications). Based on a novel set of classical multiparticle solutions, we propose a robust framework to describe
scalars, gluons, gravitons, and their interactions at tree
level. Contrary to flat space, the multiparticle recursion in AdS momentum space is not algebraic, and involves the inversion of differential
operators in the radial coordinate. The key step here is a suitable
gauge choice. Instead of the traditional axial gauge, we define a
\emph{boundary transversal gauge}. While equivalent at the linearized
level, the latter lets us localize all the tensor structure into the vertices, exclusively working with scalar propagators. 

The multiparticle currents are given by nested integrals in the radial coordinate and can be used to compute $N$-point tree-level boundary correlators. Because of the boundary, the usual BG prescription must be generalized to recover the permutation symmetry of the correlators. In Yang-Millls (YM), for example, this prescription makes the cyclicity of the color-ordered correlators manifest while removing unphysical boundary terms.

We start by discussing classical equations of motion in AdS, with the introduction of a convenient gauge choice for handling the multiparticle solutions. First, we look at the YM theory and the color-stripped perturbiner. Next, we analyze graviton multiparticle solutions, and finalize with the discussion of scalars coupled to YM and gravity. In each case we propose and verify the prescription for tree level correlators. Along the way we explain how to adapt our recursion to dS, where they compute coefficients of the cosmological wavefunction \cite{Maldacena:2002vr,Maldacena:2011nz,McFadden:2010vh}. We then present some final remarks and natural directions to investigate next. 

\section{Field equations in AdS}

We work with AdS$_{d+1}$ with radius $\mathcal{R}$ in the Poincar\'e patch, 
\begin{equation}
\tilde{g}_{mn}dx^{m}dx^{n}=\tfrac{\mathcal{R}^{2}}{z^{2}}(dz^{2}+\eta_{\mu\nu}dx^{\mu}dx^{\nu}),
\end{equation}
with $0 <z<\infty$. The spacetime indices $m,n,\ldots$ generically represent the radial
direction $z$ and the boundary directions. The latter are denoted
by $\mu,\nu=0,\ldots,d-1$, and $\eta_{\mu\nu}$ is the flat boundary
metric (Lorentzian). We will often use the shorthand $(U\cdot V)=\eta^{\mu\nu}U_{\mu}V_{\nu}$
for boundary vectors. For dS we take the boundary metric to be Euclidean and Wick-rotate the radial coordinate, $z\to-i\eta$.

We start with a scalar field with mass $\mathrm{m}$ coupled to the curvature $R$ via a constant parameter $\xi$, satisfying\noindent
\begin{equation}
g^{mn}\partial_{m}\partial_{n}\phi-g^{mn}\Gamma_{mn}^{p}\partial_{p}\phi=(\mathrm{m}^{2}+\xi R)\phi.\label{eq:eom-scalar}
\end{equation}
The left hand side is simply the curved d'Alembertian, with $\Gamma_{mn}^{p}=g^{pq}\Gamma_{mnq}$
denoting the Christoffel symbol\noindent
\begin{equation}
\Gamma_{mnp}[g]=\tfrac{1}{2}(\partial_{m}g_{np}+\partial_{n}g_{mp}-\partial_{p}g_{mn}).\label{eq:Christoffel}
\end{equation}
In the rest of this letter we take the free solutions to be eigenstates of the
boundary momenta, denoted by $k_{\mu}$.  In the Poincar\'e patch ($g_{mn}=\tilde{g}_{mn}$), equation \eqref{eq:eom-scalar}
is recast as
\begin{eqnarray}
\mathcal{D}_{k}^{2} \phi & = & M^{2}\phi , \label{eq:scalar-ads}\\
\mathcal{D}_{k}^{2} & \equiv & z^{2}\partial_{z}^{2}+(1-d)z\partial_{z}-z^{2}k^{2},
\end{eqnarray}
with $k^{2}=(k\cdot k)$, and effective mass  $M^{2} = (\mathrm{m}\mathcal{R})^{2}-\xi d(d+1)$. The solutions of \eqref{eq:scalar-ads} are Bessel functions (or Hankel functions for dS). Under proper boundary conditions and normalization, they are identified with (A)dS bulk-to-boundary propagators (see e.g. \cite{Albayrak:2020fyp,Gomez:2021ujt} for more details).

The curved Yang-Mills equations are given by
\begin{multline}
g^{np}\partial_{p}\mathbf{F}_{mn}=ig^{np}[\mathbf{A}_{p},\mathbf{F}_{mn}]+\mathbf{J}_{m}\\
+g^{np}(\Gamma_{mp}^{q}\mathbf{F}_{qn}+\Gamma_{np}^{q}\mathbf{F}_{mq}),\label{eq:eom-gluon}
\end{multline}
where $\mathbf{F}_{mn}=\partial_{m}\mathbf{A}_{n}-\partial_{n}\mathbf{A}_{m}-i[\mathbf{A}_{m},\mathbf{A}_{n}]$
is the field strength, $\mathbf{A}_{m}$ is Lie algebra valued for
some unspecified gauge group with generators $T^{a}$, and $\mathbf{J}_{m}$
generically denotes the coupling to other fields. We take $\mathbf{A}_{\mu}=(\mathcal{R}/z)A_{\mu}$ and $\mathbf{A}_{z}=(\mathcal{R}/z)\alpha$,
such that the linearized version of \eqref{eq:eom-gluon} is rewritten
as\begin{subequations}\label{eq:gluon-pol}
\begin{eqnarray}
(\mathcal{D}_{k}^{2}+d-1)A_{\mu} & = & izk_{\mu}[z\partial_{z}+(2-d)]\alpha\nonumber \\
 &  & -z^{2}k_{\mu}(k\cdot A),\\
k^{2}\alpha & = & i(1/z-\partial_{z})(k\cdot A).
\end{eqnarray}
\end{subequations}Instead of the axial gauge $\alpha=0$, we will
choose the boundary transversal gauge,
\begin{equation}
\eta^{\mu\nu}\partial_{\mu}A_{\nu}=0.\label{eq:BTG-gluon}
\end{equation}
They are equivalent at the linearized level: when $k^{2}\neq0$ we have $\alpha=0$, while for $k^{2}=0$ we set $\alpha$ to zero via a residual gauge symmetry.

Finally, we review Einstein's field equations with cosmological constant
$\Lambda=d(1-d)/(2\mathcal{R}^{2})$. In the presence of matter, with action $S_{\textrm{matter}}$ and
energy-momentum tensor 
\begin{equation}
T_{mn}\equiv-\tfrac{2}{\sqrt{-g}}\tfrac{\delta}{\delta g^{mn}}S_{\textrm{matter}},
\end{equation}
they can be cast as
\begin{equation}
R_{mn}+\tfrac{d}{\mathcal{R}^{2}}g_{mn}=\kappa T_{mn}-\tfrac{\kappa}{(d-1)}g_{mn}g^{pq}T_{pq},\label{eq:eom-graviton}
\end{equation}
with gravitational coupling $\kappa$, Ricci tensor $R_{mn}$ given
by
\begin{equation}
R_{mn}=\partial_{p}\Gamma_{mn}^{p}-\partial_{n}\Gamma_{mp}^{p}+\Gamma_{pq}^{p}\Gamma_{mn}^{q}-\Gamma_{nq}^{p}\Gamma_{mp}^{q},
\end{equation}
and scalar curvature $R=g^{mn}R_{mn}$.

The graviton dynamics can be accessed via a deformation of the background metric. Gravitons are
parametrized here as
\begin{equation}
g_{mn}=\tilde{g}_{mn}+\tfrac{\mathcal{R}^{2}}{z^{2}}h_{mn},\label{eq:graviton-parametrization}
\end{equation}
The analogue of the boundary transversal gauge is\begin{subequations}\label{eq:BTG-graviton}
\begin{align}
\eta^{\mu\nu}\partial_{\mu}h_{z\nu} & =\tfrac{1}{2}\eta^{\mu\nu}\partial_{z}h_{\mu\nu}+\tfrac{d}{2z}h_{zz},\\
\eta^{\nu\rho}\partial_{\rho}h_{\mu\nu} & =\tfrac{1}{2}\partial_{\mu}(\eta^{\nu\rho}h_{\nu\rho}+\beta h_{zz}),
\end{align}
\end{subequations}where $\beta$ is a constant parameter. Then the linearized
version of \eqref{eq:eom-graviton} is given by\begin{subequations}\label{eq:graviton-pol}
\begin{align}
k^{2}h_{zz} & =0,\\
k^{2}h_{z\mu} & =\tfrac{i}{2z}(d-2-\beta z\partial_{z})k_{\mu}h_{zz},\\
\mathcal{D}_{k}^{2}h_{\mu\nu} & =[(1-\beta)z^{2}k_{\mu}k_{\nu}+\eta_{\mu\nu}(d-z\partial_{z})]h_{zz}\nonumber \\
 & +[z^{2}\partial_{z}+(1-d)z](\partial_{\mu}h_{\nu z}+\partial_{\nu}h_{\mu z}).
\end{align}
\end{subequations}Like in Yang-Mills, the components $h_{zz}$
and $h_{z\mu}$ vanish on-shell ($k^{2}\neq0$) or via a residual
gauge transformation ($k^{2}=0$). 

With our gauge choice,  the physical degrees of freedom both in YM and in gravity have scalar propagators, with interesting consequences in the context of multiparticle solutions.

\section{Multigluon solutions and correlators}

We are now going to evaluate the multiparticle solutions of \eqref{eq:eom-gluon}
through the ansatz:
\begin{subequations}\label{eq:gluon-multi}
\begin{eqnarray}
\mathbf{A}_{\mu}(x,z) & = & \tfrac{\mathcal{R}}{z}\sum_{I}\mathcal{A}_{I\mu}(z)T^{a_{I}}e^{ik_{I}\cdot x},\\
\mathbf{A}_{z}(x,z) & = & \tfrac{\mathcal{R}}{z}\sum_{I}\alpha_{I}(z)T^{a_{I}}e^{ik_{I}\cdot x},\\
\mathbf{J}_{m}(x,z) & = & \sum_{I}\mathcal{J}_{Im}(z)T^{a_{I}}e^{ik_{I}\cdot x}.
\end{eqnarray}
\end{subequations}
The word $I$ denotes a sequence of letters $I=i_{1}\ldots i_{\ell}$,
where $i$ is a single-particle label, with $k_{I}\equiv k_{i_{1}}+\ldots+k_{i_{\ell}}$
and $T^{a_{I}}=T^{a_{i_{1}}}\cdots T^{a_{i_{\ell}}}$. The boundary
transversal gauge translates to $(k_{I}\cdot\mathcal{A}_{I})=0$.
The single-particle solutions of \eqref{eq:gluon-pol}, i.e. bulk-to-boundary propagators, are then associated
to one-letter words, which we denote by $\mathcal{A}_{i\mu}=\varepsilon_{i\mu}\tilde{\phi}(z)e^{ik_{i}\cdot x}$
and $\alpha_{i}=0$. The polarization $\varepsilon_{i\mu}$ is transversal
$(k_{i}\cdot\varepsilon_{i})=0$, and $\tilde{\phi}$ satisfies $(\mathcal{D}_{i}^{2}+d-1)\tilde{\phi}=0$.
We refer to $\mathcal{A}_{I\mu}$ and $\alpha_{I}$
as multiparticle currents. The specific form of $\mathbf{J}_{m}$
depends on the model, and we will see an explicit example later.

After plugging the above ansatz in \eqref{eq:eom-gluon}, we obtain the multiparticle recursions
\begin{multline}
\tfrac{1}{z^{2}}(\mathcal{D}_{I}^{2}+d-1)\mathcal{A}_{I\mu}=\vphantom{\sum_{I=JK}}ik_{I\mu}[\partial_{z}+(2-d)/z]\alpha_{I}-\tfrac{\mathcal{R}}{z}\mathcal{J}_{I\mu}\\
+\tfrac{\mathcal{R}}{z}\sum_{I=JK}\{(k_{K\mu}\alpha_{K}+2i\partial_{z}\mathcal{A}_{K\mu})\alpha_{J}+k_{K\mu}(\mathcal{A}_{J}\cdot\mathcal{A}_{K})\\
+\vphantom{\sum_{I=JK}}\mathcal{A}_{K\mu}[i(\partial_{z}-d/z)\alpha_{J}-2(k_{K}\cdot\mathcal{A}_{J})]-(J\leftrightarrow K)\}\\
+\tfrac{\mathcal{R}^{2}}{z^{2}}\sum_{I=JKL}\{[\alpha_{J}\alpha_{K}\mathcal{A}_{L\mu}+(\mathcal{A}_{J}\cdot\mathcal{A}_{K})\mathcal{A}_{L\mu}-(K\leftrightarrow L)]\\
\vphantom{\sum_{I=JK}}+[\alpha_{K}\alpha_{L}\mathcal{A}_{J\mu}+(\mathcal{A}_{K}\cdot\mathcal{A}_{L})\mathcal{A}_{J\mu}-(J\leftrightarrow K)]\},\label{eq:recursion-A}
\end{multline}
and
\begin{multline}
k_{I}^{2}\alpha_{I}=\tfrac{\mathcal{R}}{z}\sum_{I=JK}\{2\alpha_{K}(k_{K}\cdot\mathcal{A}_{J})-2\alpha_{J}(k_{J}\cdot\mathcal{A}_{K})\\
\vphantom{\sum_{I=JK}}+i(\mathcal{A}_{J}\cdot\partial_{z}\mathcal{A}_{K})-i(\mathcal{A}_{K}\cdot\partial_{z}\mathcal{A}_{J})\}+\tfrac{\mathcal{R}}{z}\mathcal{J}_{Iz}\\
+\tfrac{\mathcal{R}^{2}}{z^{2}}\sum_{I=JKL}\{\alpha_{K}(\mathcal{A}_{J}\cdot\mathcal{A}_{L})-\alpha_{L}(\mathcal{A}_{J}\cdot\mathcal{A}_{K})+(J\leftrightarrow L)\},\label{eq:recursion-alpha}
\end{multline}
with shorthand $\mathcal{D}_{I}^{2}=\mathcal{D}_{k_{I}}^{2}$. The
operation $I=JK$ ($JKL$) denotes a deconcatenation, which consists
of all the \emph{order preserving} ways of splitting the
word $I$ into $JK$ ($JKL$). Note that $\mathcal{A}_{I\mu},\alpha_I$ do not carry any color structure, which has been stripped off in \eqref{eq:gluon-multi}.

\begin{figure}
\begin{center}
\includegraphics[width=\linewidth]{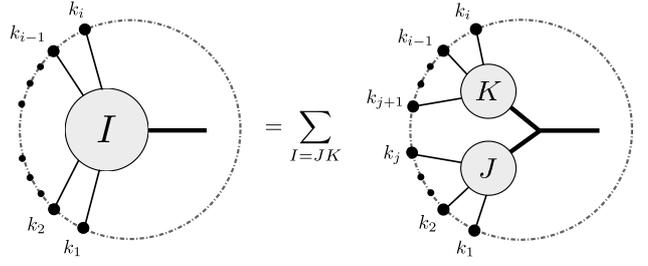}
\end{center}
\caption{Graphic representation of the cubic vertex deconcatenation $I=JK$ in the multiparticle recursion, with $I=12...i$. The dashed line denotes the AdS boundary. Thin (thick) lines denote bulk-to-boundary (bulk-to-bulk) propagators.}
\label{fig:gluon-recursion}
\end{figure}

The inversion of $(\mathcal{D}_{I}^{2}-M^{2})$
is defined via the Green function $G_{I}(z,y)$, a bulk-to-bulk propagator, satisfying
\begin{equation}
(\mathcal{D}_{I}^{2}-M^{2})G_{I}=z^{d+1}\delta(z-y)
\end{equation}
with appropriate boundary conditions. In particular, we have
\begin{equation}
(\mathcal{D}_{I}^{2}-M^{2})^{-1}\mathcal{O}(z)=\int\tfrac{dy}{y^{d+1}}G_{I}(z,y)\mathcal{O}(y),
\end{equation}
and the recursion of $A_{I\mu}$, depicted in figure \ref{fig:gluon-recursion},
is computed through nested integrals in the radial coordinate. Explicit expressions for $G_I$ in AdS were derived in \cite{Raju:2011mp}. Wick rotating them to dS is subtle, see \cite{Sleight:2021plv,Meltzer:2021zin,Gomez:2021ujt} for further details.

The prescription for computing $N$-gluon \textit{color-ordered} correlators is defined to be
\begin{multline}
A(1,\ldots,N)=-\frac{1}{N}\int\tfrac{dz}{z^{d+1}}\eta^{\mu\nu}\times\\
[\mathcal{A}_{1\mu}(\mathcal{D}_{2\ldots N}^{2}+d-1)\mathcal{A}_{2\ldots N\nu}+\textrm{cyc}(1,\ldots,N)],\label{eq:n-point-gluon}
\end{multline}
where boundary momentum conservation is implicit. Equation \eqref{eq:n-point-gluon} effectively removes the bulk-to-bulk propagator from $\mathcal{A}_{2\ldots N\nu}$ (right-most one in Figure \ref{fig:gluon-recursion}), and replaces it by a bulk-to-boundary propagator. This is a straightforward generalization of the usual Berends-Giele prescription \cite{Berends:1987me}. The extra ingredients here are the integration over the radial coordinate and the explicit sum over the cyclic permutations of the $N$ external legs, which is redundant in flat space. The latter removes unphysical boundary contributions that would otherwise break the cyclicity of the color-ordered correlators.

Let us now present a couple of examples. The three-point
result is given by
\begin{equation}
A(1,2,3)=\mathcal{R}^{3}\varepsilon_{1\mu}\varepsilon_{2\nu}\varepsilon_{3\rho}V_{123}^{\mu\nu\rho}\int\tfrac{dz}{z^{d}}\tilde{\phi}_{1}\tilde{\phi}_{2}\tilde{\phi}_{3},
\end{equation}
with the usual polarization structure of Yang-Mills,
\begin{equation}
V_{123}^{\mu\nu\rho}=\eta^{\mu\nu}\eta^{\rho\sigma}(k_{1}-k_{2})_{\sigma}+\textrm{cyc}(1\mu,2\nu,3\rho).\label{eq:YM-vertex}
\end{equation}
The four-point correlator is given by
\begin{multline}
A(1,2,3,4)=\Pi_{12|34}^{0}\int\tfrac{dz}{z^{d+1}}(\tilde{\phi}_{1}\overleftrightarrow{\partial_{z}}\tilde{\phi}_{2})(\tilde{\phi}_{3}\overleftrightarrow{\partial_{z}}\tilde{\phi}_{4})\\
+\Pi_{12|34}^{1}\int\tfrac{dz}{z^{d+1}}(z\tilde{\phi}_{1}\tilde{\phi}_{2})(\mathcal{D}_{34}^{2}+d-1)^{-1}(z\tilde{\phi}_{3}\tilde{\phi}_{4})\\{}
+[(\varepsilon_{1}\cdot\varepsilon_{3})(\varepsilon_{2}\cdot\varepsilon_{4})-(3\leftrightarrow4)]\int\tfrac{dz}{z^{d+1}}\tilde{\phi}_{1}\tilde{\phi}_{2}\tilde{\phi}_{3}\tilde{\phi}_{4}\\
-[(34)\to(23)],\label{eq:4pt-gluon}
\end{multline}
with $U\overleftrightarrow{\partial_{z}}V=U\partial_{z}V-V\partial_{z}U$.
In the first line the polarization structure is encoded in
\begin{equation}
\Pi_{12|34}^{0}=\tfrac{\mathcal{R}^{4}}{k_{34}^{2}}(\varepsilon_{1}\cdot\varepsilon_{2})(\varepsilon_{3}\cdot\varepsilon_{4}),
\end{equation}
while in the second line we have
\begin{multline}
\Pi_{12|34}^{1}=\mathcal{R}^{4}\tfrac{(k_{1}^{2}-k_{2}^{2})(k_{3}^{2}-k_{4}^{2})}{k_{34}^{2}}(\varepsilon_{1}\cdot\varepsilon_{2})(\varepsilon_{3}\cdot\varepsilon_{4})\\
\vphantom{\tfrac{(k_{1}^{2}-k_{2}^{2})(k_{3}^{2}-k_{4}^{2})}{k_{34}^{2}}}+\mathcal{R}^{4}\eta^{\mu\nu}[2\varepsilon_{1\mu}(k_{1}\cdot\varepsilon_{2})-k_{1\mu}(\varepsilon_{1}\cdot\varepsilon_{2})-(1\leftrightarrow2)]\times\\
\vphantom{\tfrac{(k_{1}^{2}-k_{2}^{2})(k_{3}^{2}-k_{4}^{2})}{k_{34}^{2}}}[2\varepsilon_{3\nu}(k_{3}\cdot\varepsilon_{4})-k_{3\nu}(\varepsilon_{3}\cdot\varepsilon_{4})-(3\leftrightarrow4)].
\end{multline}
The third line in \eqref{eq:4pt-gluon} is simply a four-point contact Witten diagram.

Because of the boundary transversal gauge, the correlators
computed via \eqref{eq:n-point-gluon} are expressed in terms
of scalar-like factorization channels. The price to pay is the apparent
introduction of spurious poles of the form $k_{ij}^{-2}$. The final
expression, however, is equivalent to other results in the literature.
For example, we match \eqref{eq:4pt-gluon} with the results of \cite{Baumann:2020dch} when $d=3$.

\section{Multigraviton solutions and correlators}

For the multiparticle solutions of \eqref{eq:eom-graviton}, we start
with an ansatz inspired by the parametrization \eqref{eq:graviton-parametrization},
\begin{equation}
g_{mn}=\tilde{g}_{mn}+\tfrac{\mathcal{R}^{2}}{z^{2}}\sum_{I}\mathcal{H}_{Imn}e^{ik_{I}\cdot x}.\label{eq:ansatz-H}
\end{equation}
The main difference with YM is the absence of the
color structure, so we consider only the sum over ordered words $I=i_{1}\ldots i_{\ell}$,
with $i_{1}<i_{2}<\ldots<i_{\ell}$. 

The natural multiparticle ansatz for $g^{mn}$ is
\begin{equation}
g^{mn}=\tilde{g}^{mn}-\tfrac{\mathcal{R}^{2}}{z^{2}}\sum_{I}\mathcal{I}_{I}^{mn}e^{ik_{I}\cdot x}.\label{eq:ansatz-I}
\end{equation}
Since the inverse metric satisfies $g^{mp}g_{np}=\delta_{n}^{m}$, the multiparticle currents in \eqref{eq:ansatz-I} are constrained to be
\begin{equation}
\mathcal{I}_{I}^{mn}=\tilde{g}^{mp}\mathcal{H}_{Ipq}\tilde{g}^{qn}-\tfrac{\mathcal{R}^{2}}{z^{2}} \sum_{I=J\cup K}\mathcal{I}_{J}^{mp}\mathcal{H}_{Kpq} \tilde{g}^{qn}.\label{eq:g-inverse}
\end{equation}
The operation $I=J\cup K$ denotes the deshuffle, which means we consider
all possible ways of splitting the ordered word $I$ into two non-empty
ordered words $J$ and $K$. Equation \eqref{eq:g-inverse} is responsible
for packing the infinite number of vertices in gravity into a simple
recursion \cite{Gomez:2021shh}. In practice, the recursive structure is encoded in up to quintic interaction vertices, which is a vast improvement over standard diagrammatic techniques.

In terms of the multiparticle currents, gauge \eqref{eq:BTG-graviton}
reads\begin{subequations}
\begin{align}
i\eta^{\mu\nu}k_{I\mu}\mathcal{H}_{Iz\nu} & =\tfrac{1}{2}\eta^{\mu\nu}\partial_{z}\mathcal{H}_{I\mu\nu}+\tfrac{d}{2z}\mathcal{H}_{Izz},\\
i\eta^{\nu\rho}k_{I\rho}\mathcal{H}_{I\mu\nu} & =\tfrac{i}{2}k_{I\mu}(\eta^{\nu\rho}\mathcal{H}_{I\nu\rho}+\beta\mathcal{H}_{Izz}),
\end{align}
\end{subequations}and the ansatz \eqref{eq:ansatz-H}
solves equation \eqref{eq:eom-graviton} when the multiparticles currents
satisfy\begin{subequations}\label{eq:recursion-H}
\begin{equation}
k_{I}^{2}\mathcal{H}_{Izz}=\tfrac{2\kappa}{(d-1)}[(d-2)\mathcal{T}_{Izz}-\eta^{\mu\nu}\mathcal{T}_{I\mu\nu}]-2\mathcal{G}_{Izz},
\end{equation}
\begin{multline}
k_{I}^{2}\mathcal{H}_{Iz\mu}=2\kappa\mathcal{T}_{Iz\mu}-2\mathcal{G}_{Iz\mu}\\
+\tfrac{i}{2z}(d-2-\beta z\partial_{z})k_{I\mu}\mathcal{H}_{Izz},
\end{multline}
\begin{multline}
\mathcal{D}_{I}^{2}\mathcal{H}_{I\mu\nu}=\tfrac{2\kappa z^{2}}{(d-1)}\eta_{\mu\nu}(\mathcal{T}_{Izz}+\eta^{\rho\sigma}\mathcal{T}_{I\rho\sigma})-2\kappa z^{2}\mathcal{T}_{I\mu\nu}\\
+\vphantom{\tfrac{2\kappa z^{2}}{(d-1)}}[(1-\beta)z^{2}k_{I\mu}k_{I\nu}+\eta_{\mu\nu}(d-z\partial_{z})]\mathcal{H}_{Izz}+2z^{2}\mathcal{G}_{I\mu\nu}\\
+\vphantom{\tfrac{2\kappa z^{2}}{(d-1)}}iz[z\partial_{z}+(1-d)](k_{I\mu}\mathcal{H}_{Iz\nu}+k_{I\nu}\mathcal{H}_{Iz\mu}).
\end{multline}
\end{subequations}
$\mathcal{T}_{Imn}$ denotes the currents of the
multiparticle expansion of the energy-momentum tensor, $T_{mn}=\sum_{I}\mathcal{T}_{Imn}e^{ik_{I}\cdot x}$. The interaction between gravitons and matter in AdS is captured by $\mathcal{G}_{Imn}$, which is fully displayed in Appendix \ref{sec:gravity-interactions}. By construction,
the currents $\mathcal{H}_{Imn}$ are symmetric under the permutation
of any single-particle labels. The single-particle solutions of \eqref{eq:graviton-pol}
are again associated to one-letter words, which we denote by $\mathcal{H}_{i\mu\nu}=h_{i\mu\nu}\varphi(z)e^{ik_{i}\cdot x}$
and $\mathcal{H}_{iz\mu}=\mathcal{H}_{izz}=0$.
The boundary polarization $h_{i\mu\nu}$ is traceless ($\eta^{\mu\nu}h_{i\mu\nu}=0$)
and transversal ($\eta^{\nu\rho}k_{i\rho}h_{i\mu\nu}=0$), and $\varphi(z)$ is a massless minimally coupled ($\xi=0$) scalar.

Like in YM, the recursions in \eqref{eq:recursion-H}
present a characteristic feature of the boundary transversal gauge: the tensor structure of the correlator is relegated to the interaction vertices and only scalar propagators appear. Morevoer the currents $\mathcal{H}_{Izz}$, $\mathcal{H}_{Iz\mu}$, as well as $\eta^{\mu\nu}\mathcal{H}_{I\mu\nu}$ and $\eta^{\nu\rho}k_{I\rho}\mathcal{H}_{I\mu\nu}$, have a trivial propagator.

The generalization of the color-ordered correlators in \eqref{eq:n-point-gluon} to gravity is given by
\begin{multline}
\mathcal{M}_{N}=-\frac{1}{N}\kappa\int\tfrac{dz}{z^{d+1}}\eta^{\mu\rho}\eta^{\nu\sigma} \mathcal{H}_{1\mu\nu}(\mathcal{D}_{2\ldots N}^{2}\mathcal{H}_{2\ldots N\rho\sigma})\\
+\textrm{perm}(1\to2\ldots N).\label{eq:n-graviton-amplitude}
\end{multline}
The permutation in the last line makes the correlator manifestly symmetric in all $N$ legs.

Since graviton correlators quickly grow in size, we present
explicitly only the three-point case:
\begin{multline}
\mathcal{M}_{3}=\frac{\kappa}{4} h_{1\mu\nu}h_{2\rho\sigma}h_{3\gamma\lambda}\big\{ V_{123}^{\mu\rho\gamma}V_{123}^{\nu\sigma\lambda}\int\tfrac{dz}{z^{d-1}}\varphi_{1}\varphi_{2}\varphi_{3}\\
-\frac{1}{3}\eta^{\nu\rho}\eta^{\sigma\gamma}\eta^{\lambda\mu}\int dz\partial_{z}[\tfrac{1}{z^{d-1}}\partial_{z}(\varphi_{1}\varphi_{2}\varphi_{3})]\big\}.\label{eq:3pt-graviton}
\end{multline}
The first line is the well-known expression in terms of the cubic
vertices of Yang-Mills \eqref{eq:YM-vertex}. The second line encodes contact terms which have delta function support when Fourier-transformed to position space. Therefore it vanishes for generic boundary positions of the operators. In momentum space, they are characterized by being analytic in at least two of the momenta \cite{Maldacena:2011nz}. The total derivative in \eqref{eq:3pt-graviton} diverges when $z \to 0$, so we introduce a cutoff at $z=\epsilon$. After dropping the power-law divergent pieces, we find that
\begin{equation}
\mathcal{M}_{\textrm{B}}^d \equiv \int dz\partial_{z}[\tfrac{1}{z^{d-1}}\partial_{z}(\varphi_{1}\varphi_{2}\varphi_{3})]\propto\sum_{i=1}^{3}k_{i}^{d},
\end{equation}
in odd $d$, which can be removed by a redefinition of the bulk metric \cite{Maldacena:2011nz}.
For even $d$, we obtain
\begin{equation}
\mathcal{M}_{\textrm{B}}^d \propto \sum_{i=1}^{3}k_{i}^{d}\ln(\tfrac{1}{2}\epsilon k_{i}e^{\gamma_{E}})+...,
\end{equation}
where the first term can also be removed by a redefinition of the metric \cite{paulm} and the ellipsis denote polynomials in the squares of momenta, known as ultralocal terms \cite{Bzowski:2017poo}. 

\section{Scalars, gluons, and gravitons}

Now we turn our attention to scalar theories. Since their classical
multiparticle solutions have a very simple structure, we will focus
on the more interesting cases with coupling to gluons and gravitons. 

Consider first scalars in the adjoint representation of the gauge
group. Their color-stripped multiparticle expansion is analogous
to \eqref{eq:gluon-multi}, given by
\begin{equation}
\phi=\sum_{I}\Phi_{I}(z)T^{a_{I}}e^{ik_{I}\cdot x}.
\end{equation}
Single-particle states satisfy $(\mathcal{D}_{i}^{2}-M^{2})\Phi_{i}=0$,
and we consider a minimal coupling with gluons,
\begin{equation}
\partial_{m}\phi\to\partial_{m}\phi-i[\mathbf{A}_{m},\phi],
\end{equation}
such that $\mathbf{J}_{m}=[(i\partial_{m}\phi+[\mathbf{A}_{m},\phi]),\phi]$
in \eqref{eq:eom-gluon}.

In the gauge \eqref{eq:BTG-gluon}, equation \eqref{eq:eom-scalar}
minimally coupled to YM yields the following recursion,
\begin{multline}
\tfrac{1}{z^{2}}(\mathcal{D}_{I}^{2}-M^{2})\Phi_{I}=\tfrac{\mathcal{R}}{z}\sum_{I=JK}[2\Phi_{J}(k_{J}\cdot\mathcal{A}_{K})\\
\vphantom{\sum_{I=JK}}-i(\Phi_{J}\partial_{z}\alpha_{K}+2\alpha_{K}\partial_{z}\Phi_{J}-\tfrac{d}{z}\Phi_{J}\alpha_{K})-(J\leftrightarrow K)]\\
+\tfrac{\mathcal{R}^{2}}{z^{2}}\sum_{I=JKL}[(\mathcal{A}_{J}\cdot\mathcal{A}_{K})\Phi_{L}-(\mathcal{A}_{J}\cdot\mathcal{A}_{L})\Phi_{K}\\
\vphantom{\sum_{I=JK}}+\alpha_{J}\alpha_{K}\Phi_{L}-\alpha_{J}\alpha_{L}\Phi_{K}+(J\leftrightarrow L)],
\end{multline}
with color-ordered $N$-point correlators defined via
\begin{multline}
A(1,\ldots,N)=-\frac{1}{N}\int\tfrac{dz}{z^{d+1}}\Phi_{1}(\mathcal{D}_{2\ldots N}^{2}-M^{2})\Phi_{2\ldots N}\\
+\textrm{cyc}(1,\ldots,N).
\end{multline}

As an example, we take the case of four external scalars exchanging
gluons:
\begin{multline}\label{eq:4pt-scalar-gluon}
A(1,2,3,4)=\tfrac{\mathcal{R}}{2}\int\tfrac{dz}{z^{d}}\{\Phi_{1}\Phi_{2}[(k_{1}-k_{2})\cdot\mathcal{A}_{34}]\\
+i(\Phi_{1}\partial_{z}\Phi_{2}-\Phi_{2}\partial_{z}\Phi_{1})\alpha_{34}\}+\textrm{cyc}(1,2,3,4).
\end{multline}
For a conformally coupled scalar ($M^2=1-d$), this expression can be directly obtained from the YM result in \eqref{eq:4pt-gluon} with the identifications $\tilde{\phi}_{i}\to\Phi_{i}$,
$(k_{i}\cdot\varepsilon_{j})\to0$, and $(\varepsilon_{i}\cdot\varepsilon_{j})\to1$. The final result matches the form obtained in \cite{Baumann:2020dch} for $d=3$.

When graviton excitations are considered, the color structure cannot
be stripped off from the multiparticle currents, which would explicitly
involve color indices. For simplicity we will turn off the gluons
and consider a colorless scalar
\begin{equation}
\phi=\sum_{I}\Phi_{I}(z)e^{ik_{I}\cdot x},
\end{equation}
as the multiparticle ansatz solving equation \eqref{eq:eom-scalar}.
We then obtain the recursion for $\Phi_{I}$,
\begin{multline}
\tfrac{1}{z^{2}}(\mathcal{D}_{I}^{2}-M^{2})\Phi_{I}=\tfrac{\mathcal{R}^{4}}{z^{4}}\sum_{I=J\cup K}\{\mathcal{I}_{J}^{mn}\partial_{m}\partial_{n}\Phi_{K}\\
\vphantom{\sum_{I=J\cup K}}+\tfrac{z^{2}}{\mathcal{R}^{2}}\tilde{g}^{mn}[\tilde{g}^{pq}\Gamma_{Jmnq}\partial_{p}\Phi_{K}-\tfrac{2\xi}{(d-1)}\kappa\Phi_{J}\mathcal{T}_{Kmn}]\\
\vphantom{\sum_{I=J\cup K}}-\tilde{\Gamma}_{mnq}(\tilde{g}^{mn}\mathcal{I}_{J}^{pq}+\tilde{g}^{pq}\mathcal{I}_{J}^{mn})\partial_{p}\Phi_{K}\}+\ldots.\label{eq:scalar-graviton}
\end{multline}
The ellipsis denotes contributions with higher order deshuffles,
which are spelled out in Appendix \ref{sec:scalar-gravitons}.
The current $\Gamma_{Imnp}$ is defined through \eqref{eq:Christoffel}
as
\begin{equation}
\Gamma_{Imnp}=\Gamma_{mnp}[\tfrac{\mathcal{R}^{2}}{z^{2}}\mathcal{H}_{I}].
\end{equation}
The notation $\partial_{p}\mathcal{O}_{I}=i\delta_{p}^{\mu}k_{I\mu}\mathcal{O}_{I}+\delta_{p}^{z}\partial_{z}\mathcal{O}_{I}$
is implicit for any current $\mathcal{O}_{I}$. Finally, $\mathcal{T}_{Imn}$
denotes the multiparticle coefficients of the energy-momentum tensor:
\begin{multline}
T_{mn}=\partial_{m}\phi\partial_{n}\phi-\tfrac{1}{2}g_{mn}(g^{pq}\partial_{p}\phi\partial_{q}\phi+m^{2}\phi^{2})\\
+\xi(R_{mn}-\tfrac{1}{2}g_{mn}R)\phi^{2}+\xi(g_{mn}g^{pq}\partial_{p}\partial_{q}-\partial_{m}\partial_{n})\phi^{2}\\
-\xi(g_{mn}g^{pq}g^{rs}\Gamma_{pqr}\partial_{s}-g^{pq}\Gamma_{mnp}\partial_{q})\phi^{2}.
\end{multline}

The $N$-point scalar correlator is given by
\begin{multline}\label{eq:A_nscalarsG}
A_{N}=-\frac{1}{N}\int\tfrac{dz}{z^{d+1}}\Phi_{1}(\mathcal{D}_{2\ldots N}^{2}-M^{2})\Phi_{2\ldots N}\\
+\textrm{perm}(1\to2\ldots N).
\end{multline}
We have explicitly checked that the four-point correlator matches the Witten diagram calculation modulo gauge-dependent contact terms for the case $M^2=0$ (see appendix \ref{sec:scalar-gravitons}). This case is of particular interest since it arises from the dimensional reduction of the 4-point graviton amplitude in the flat space limit \cite{Herderschee:2022ntr}. Note that the $\beta$-dependent piece coming from the gauge choice \eqref{eq:BTG-graviton} may be cast as a total derivative, 
\begin{multline}
A_{4}|_{\beta}\propto\sum_{234=ij\cup k}\int dz\partial_{z}\big\{ z^{1-d}\Phi_{1}\partial_{z}\mathcal{H}_{ijzz}\Phi_{k}\\
+\frac{(k_{k}^{2}-k_{1}^{2})}{k_{ij}^{2}}z^{1-d}[\mathcal{H}_{ijzz}(\Phi_{1}\partial_{z}\Phi_{k}-\Phi_{k}\partial_{z}\Phi_{1})]\big\}\\
+\textrm{perm}(1\to234).
\end{multline}
Once again, these boundary contributions correspond to contact terms with delta function support in position space.

\section{Final remarks}

Inspired by the perturbiner method in flat space, we have derived the first classical multiparticle solutions for scalars, gluons, and gravitons in AdS$_{d+1}$. Their recursive character requires nested integrations
in the radial coordinate, with bulk-to-bulk propagator insertions. Perhaps
more noteworthy is the fact that in any of these theories we require
only \emph{scalar} bulk-to-bulk propagators. This follows from a special
gauge choice, dubbed here boundary transversal gauge (see \eqref{eq:BTG-gluon}
for YM and \eqref{eq:BTG-graviton} for Einstein gravity). At the linear
level, it is equivalent to the axial gauge. At the non-linear level,
however, the latter makes the perturbiner recursion impractical,
introducing further differential operators in the radial coordinate. 

Our recursive approach is equivalent to the Witten diagrammatic expansion in AdS momentum space up to contact terms with delta function support when Fourier-transformed to position space. In general, Witten diagrams capture the transverse traceless part of the dual CFT correlators.  Ward identities can then be used to determine the remaining terms. They correspond to contact terms in position space and vanish for generic locations of the CFT operators \cite{Maldacena:2011nz,Bzowski:2013sza,Armstrong:2020woi}. 

Due to the (A)dS boundary, the usual flat space BG prescription had to be generalized. For Yang-Mills theory, we introduced a prescription that makes the cyclicity of color-ordered correlators manifest. We verified up to five points that this removes unphysical boundary contributions. For gravity, the prescription restores permutation invariance of the correlators. Finally, we analyzed scalars exchanging gluons and gravitons, obtaining novel formulae which we matched against four-point Witten diagrams. They exhibit interesting new structures related to the double copy \cite{Bern:2008qj,Bern:2010ue} and will be presented in \cite{AGJLM:2022}. We expect our framework to be more transparent to the color-kinematics duality, much in the same way that the flat space perturbiner could realize a BCJ gauge through a multiparticle gauge choice \cite{Lee:2015upy}.

In summary, we have established an elegant tool for computing tree-level boundary correlators in (A)dS. Our results also provide a systematic construction of higher-point graviton correlators which is currently very challenging using Witten diagrams. Exploring whether our approach exposes some hidden structures in these correlators is therefore an important priority for future work. We  plan to investigate the implications of our recursions for cosmology and the relation to other recent approaches based on the double-copy \cite{Farrow:2018yni,Lipstein:2019mpu,Armstrong:2020woi,Albayrak:2020fyp,Alday:2021odx,Jain:2021qcl,Zhou:2021gnu,Diwakar:2021juk,Sivaramakrishnan:2021srm,Cheung:2022pdk,Herderschee:2022ntr,Drummond:2022dxd,Alday:2022lkk,Armstrong:2022csc,Bissi:2022wuh}, factorization \cite{Arkani-Hamed:2015bza,Arkani-Hamed:2018kmz,Baumann:2019oyu,Baumann:2020dch, Baumann:2021fxj}, unitarity \cite{Goodhew:2020hob,Jazayeri:2021fvk,Melville:2021lst,Goodhew:2021oqg,Meltzer:2021zin}, Mellin space \cite{Sleight:2019hfp,Sleight:2021plv}, Witten diagrams \cite{Heckelbacher:2022hbq,Bzowski:2022rlz}, scattering equations  in (A)dS \cite{Eberhardt:2020ewh,Roehrig:2020kck,Gomez:2021qfd,Gomez:2021ujt}, and geometric approaches \cite{Arkani-Hamed:2017fdk}.

One of the claims to fame of BG recursion in flat space is the first proof of the Parke-Taylor formula \cite{Parke:1986gb} for all tree-level MHV amplitudes in YM \cite{Berends:1987me}. In four-dimensional (A)dS, the natural analogue of MHV amplitudes are tree-level all-plus correlators of gluons, which vanish in the flat space limit. It would be truly rewarding if the recursion relations we formulate in this paper could suggest all mutliplicity formulae for such correlators.

\begin{acknowledgments}
We would like to thank Cristhiam Lopez-Arcos, Paul McFadden, Velayudhan Parameswaran Nair, Fei Teng, Alexander Quintero Velez, and Xinan Zhou for valuable discussions, comments on the draft, and reference suggestions. CA, HG, and AL are supported by the Royal Society via a PhD studentship, PDRA
grant, and a University Research Fellowship, respectively. JM is supported by a Durham-CSC Scholarship.
\end{acknowledgments}

\pagebreak
\begin{widetext}
\appendix

\section{Explicit expression of $\mathcal{G}_{Imn}$\label{sec:gravity-interactions}}

Here we spell out the multiparticle current $\mathcal{G}_{Imn}$ capturing
the interaction vertices of gravitons. This information is encoded
in the equation of motion \eqref{eq:eom-graviton} and made explicit
by the multiparticle ansatz \eqref{eq:ansatz-H}. It can be cast as
\begin{align}
\mathcal{G}_{Imn} & =\sum_{I=J\cup K}\Big\{\partial_{n}(\tfrac{\mathcal{R}^{2}}{z^{2}}\mathcal{I}_{J}^{pq}\Gamma_{Kmpq}-\tfrac{\mathcal{R}^{4}}{z^{4}}\tilde{g}^{qs}\mathcal{I}_{J}^{pr}\mathcal{H}_{Krs}\tilde{\Gamma}_{mpq})-\partial_{p}(\tfrac{\mathcal{R}^{2}}{z^{2}}\mathcal{I}_{J}^{pq}\Gamma_{Kmnq}-\tfrac{\mathcal{R}^{4}}{z^{4}}\tilde{g}^{qs}\mathcal{I}_{J}^{pr}\mathcal{H}_{Krs}\tilde{\Gamma}_{mnq})\nonumber \\
\vphantom{\sum_{I=J\cup K}} & +\tilde{g}^{pq}\tilde{g}^{rs}(\Gamma_{Jprq}\Gamma_{Kmns}-\Gamma_{Jnsq}\Gamma_{Kmpr})+\tfrac{\kappa}{(d-1)}(\tilde{g}^{pq}\tfrac{\mathcal{R}^{2}}{z^{2}}\mathcal{H}_{Jmn}\mathcal{T}_{Kpq}-\tilde{g}_{mn}\tfrac{z^{2}}{\mathcal{R}^{2}}\mathcal{I}_{J}^{pq}\mathcal{T}_{Kpq})\nonumber \\
\vphantom{\sum_{I=J\cup K}} & +\tfrac{\mathcal{R}^{2}}{z^{2}}(\tilde{g}^{pq}\mathcal{I}_{J}^{rs}+\tilde{g}^{rs}\mathcal{I}_{J}^{pq})(\tilde{\Gamma}_{nsq}\Gamma_{Kmpr}+\tilde{\Gamma}_{mpr}\Gamma_{Knsq}-\tilde{\Gamma}_{mns}\Gamma_{Kprq}-\tilde{\Gamma}_{prq}\Gamma_{Kmns})\nonumber \\
\vphantom{\sum_{I=J\cup K}} & +\tfrac{\mathcal{R}^{4}}{z^{4}}(\tilde{g}^{pq}\tilde{g}^{su}\mathcal{I}_{J}^{rt}\mathcal{H}_{Ktu}+\tilde{g}^{rs}\tilde{g}^{qu}\mathcal{I}_{J}^{pt}\mathcal{H}_{Ktu}+\mathcal{I}_{J}^{pq}\mathcal{I}_{K}^{rs})(\tilde{\Gamma}_{prq}\tilde{\Gamma}_{mns}-\tilde{\Gamma}_{mpr}\tilde{\Gamma}_{nsq})\nonumber \\
\vphantom{\sum_{I=J\cup K}} & +\sum_{I=J\cup K\cup L}\Big\{\tfrac{\mathcal{R}^{4}}{z^{4}}\mathcal{I}_{J}^{pq}\mathcal{I}_{K}^{rs}(\tilde{\Gamma}_{prq}\Gamma_{Lmns}+\Gamma_{Lprq}\tilde{\Gamma}_{mns}-\tilde{\Gamma}_{mpr}\Gamma_{Lnsq}-\Gamma_{Lmpr}\tilde{\Gamma}_{snq})\nonumber \\
\vphantom{\sum_{I=J\cup K}} & +\tfrac{\mathcal{R}^{2}}{z^{2}}\tilde{g}^{pq}\mathcal{I}_{J}^{rs}(\Gamma_{Kmpr}\Gamma_{Lnsq}+\Gamma_{Kmrp}\Gamma_{Lnqs}-\Gamma_{Kprq}\Gamma_{Lmns}-\Gamma_{Krps}\Gamma_{Lmnq})-\tfrac{\kappa}{(d-1)}\tfrac{\mathcal{R}^{4}}{z^{4}}\mathcal{H}_{Jmn}\mathcal{I}_{K}^{pq}\mathcal{T}_{Lpq}\Big\}\nonumber \\
\vphantom{\sum_{I=J\cup K}} & +\sum_{I=J\cup K\cup L\cup M}\tfrac{\mathcal{R}^{4}}{z^{4}}\mathcal{I}_{J}^{pq}\mathcal{I}_{K}^{rs}(\Gamma_{Lprq}\Gamma_{Mmns}-\Gamma_{Lnrq}\Gamma_{Mmps}).
\end{align}
The tilded variables are simply the AdS metric and derived quantities,
with nonzero components:
\begin{equation}
\begin{array}{cccc}
\tilde{g}_{zz}=\tfrac{\mathcal{R}^{2}}{z^{2}}, & \tilde{g}^{zz}=\tfrac{z^{2}}{\mathcal{R}^{2}}, & \tilde{g}_{\mu\nu}=\tfrac{\mathcal{R}^{2}}{z^{2}}\eta_{\mu\nu}, & \tilde{g}^{\mu\nu}=\tfrac{z^{2}}{\mathcal{R}^{2}}\eta^{\mu\nu},\\
\tilde{\Gamma}_{zzz}=-\tfrac{\mathcal{R}^{2}}{z^{3}}, & \tilde{\Gamma}_{\mu\nu z}=\tfrac{\mathcal{R}^{2}}{z^{3}}\eta_{\mu\nu}, & \tilde{\Gamma}_{\mu z\nu}=-\tfrac{\mathcal{R}^{2}}{z^{3}}\eta_{\mu\nu}, & \tilde{\Gamma}_{z\mu\nu}=-\tfrac{\mathcal{R}^{2}}{z^{3}}\eta_{\mu\nu}.
\end{array}
\end{equation}

\section{Scalars exchanging gravitons\label{sec:scalar-gravitons}}

For completeness, we present here the full recursion for the scalar
multiparticle currents of \eqref{eq:scalar-graviton}. It may be cast
as
as
\begin{equation}
\tfrac{1}{z^{2}}(\mathcal{D}_{I}^{2}-M^{2})\Phi_{I}=\tfrac{\mathcal{R}^{4}}{z^{4}}\sum_{k=2}^{4}\Phi_{I}^{(k)},
\end{equation}
with
\begin{eqnarray}
\Phi_{I}^{(2)} & \equiv & \sum_{I=J\cup K}[\mathcal{I}_{J}^{mn}\partial_{m}\partial_{n}\Phi_{K}+\tfrac{z^{2}}{\mathcal{R}^{2}}\tilde{g}^{mn}\tilde{g}^{pq}\Gamma_{Jmnq}\partial_{p}\Phi_{K}-\tilde{\Gamma}_{mnq}(\tilde{g}^{mn}\mathcal{I}_{J}^{pq}+\tilde{g}^{pq}\mathcal{I}_{J}^{mn})\partial_{p}\Phi_{K}\nonumber \\
 &  & -\tfrac{2\xi\kappa}{(d-1)}\tfrac{z^{2}}{\mathcal{R}^{2}}\tilde{g}^{mn}\Phi_{J}\mathcal{T}_{Kmn}],\\
\Phi_{I}^{(3)} & \equiv & \sum_{I=J\cup K\cup L}[\tfrac{\mathcal{R}^{2}}{z^{2}}\tilde{\Gamma}_{mnq}\mathcal{I}_{J}^{mn}\mathcal{I}_{K}^{pq}\partial_{p}\Phi_{L}-(\tilde{g}^{mn}\mathcal{I}_{J}^{pq}+\tilde{g}^{pq}\mathcal{I}_{J}^{mn})\Gamma_{Kmnq}\partial_{p}\Phi_{L}+\tfrac{2\xi\kappa}{(d-1)}\Phi_{J}\mathcal{I}_{K}^{mn}\mathcal{T}_{Lmn}],\\
\Phi_{I}^{(4)} & \equiv & \sum_{I=J\cup K\cup L\cup M}\tfrac{\mathcal{R}^{2}}{z^{2}}\mathcal{I}_{J}^{mn}\mathcal{I}_{K}^{pq}\Gamma_{Lmnq}\partial_{p}\Phi_{M}.
\end{eqnarray}

Following the discussion after equation \eqref{eq:A_nscalarsG}, the four-point scalar correlator exchanging gravitons is displayed below. In the gauge $\beta = 0$ with minimally coupled scalars $\xi=0$, it can be written in terms of the two-particle graviton currents as
\begin{multline}\label{eq:4pts_scalarG}
A_4= \sum_{234=ij\cup k}\int \frac{dz}{z^{d+1}} \big\{ z^2 (k_1\cdot H_{ij} \cdot k_k) \Phi_1 \Phi_k 
+\frac{z^2}{2} (k_{ij} \cdot k_k)(\eta^{\mu\nu} H_{\mu\nu ij}+ H_{zz ij}) \Phi_1 \Phi_k \\
\vphantom{\sum_{234=ij\cup k}\int_{0}^{\infty}\frac{dz}{z^{d+1}}}-iz^2 \eta^{\mu \nu} k_{\mu  k} H_{z\nu ij} (\partial_z\Phi_1 \Phi_k -\Phi_1  \partial_z\Phi_k ) - \frac{z}{2}H_{zz ij} [\partial_z \Phi_1   (z \partial_z\Phi_k) - \Phi_1  \partial_z (z \partial_z\Phi_k) ] \\
\vphantom{\sum_{234=ij\cup k}\int_{0}^{\infty}\frac{dz}{z^{d+1}}}+\frac{z^{d+1}}{2}\partial_z( z^{1-d}  H_{zz ij} \Phi_1 \partial_z\Phi_k  -2i z^{1-d} \eta^{\mu \nu} k_{\mu k} H_{z\nu ij} \Phi_1 \Phi_k )
\big\} 
+\text{perm}(1 \rightarrow 234) .
\end{multline}
Notice that the last term is a boundary contribution.

\end{widetext}
\end{document}